\def\be{\begin{equation}}
\def\ee{\end{equation}}
\def\bea{\begin{eqnarray}}

\def\eea{\end{eqnarray}}

\def\d{\delta}
\def\e{\epsilon}

\def\o{\omega}

\def\s{\sigma}

\def\bd{{\bf d}}
\def\bP{{\bf P}}
\def\bp{{\bf p}}
\def\bA{{\bf A}}

\def\bk{{\bf k}}
\def\bx{{\bf x}}
\def\bX{{\bf X}}
\def\bu{{\bf u}}

\def\ha{{1\over 2}}

\newskip\humongous \humongous=0pt plus 1000pt minus 1000pt

\newif\ifdtup

\def\(#1){(\ref{#1})}

\documentstyle[aps,psfig]{revtex}

\makeatletter                    
\@addtoreset{equation}{section}  
\makeatother                     

\begin{document}

\title{ Two-Level Atom-Field Interaction: Exact Master Equations for
Non-Markovian Dynamics, Decoherence and Relaxation }
\author{Charis Anastopoulos \thanks{ca81@umail.umd.edu} and B. L. Hu \thanks{hub@physics.umd.edu}\\
{\small Department of Physics, University of Maryland,
 College Park, Maryland 20742}}
\date{\small {\it umdpp 97-129, April 9, 1999}}
\maketitle

\begin{abstract}
We perform a first- principles derivation of the general master equation
to study the non-Markovian dynamics   of a two-level atom (2LA)
 interacting with an electromagnetic field (EMF).
We use the influence functional method which can incorporate 
the full backreaction of the field on the atom, while adopting 
Grassmannian variables for the 2LA and the coherent state representation for
the EMF. We find exact master equations for the cases of a free 
quantum field and  a cavity field in the vacuum. 
In response to the search for mechanisms to
preserve maximal coherence in quantum computations in ion trap prototypes,
we apply  these equations to analyse the decoherence of a 2LA in an EMF,
and fine that decoherence time is close to relaxation time. This 
is at variance to the claims by authors who studied the same system but
used a different coupling model. We explain the source of difference and 
argue that, contrary to common belief, the EMF when resonantly coupled 
to an atom does not decohere it as efficiently as  a bath does on a quantum 
Brownian particle. The master-equations for non-Markovian 
dynamics derived here is expected to be useful for exploring new regimes of 2LA-EMF 
interaction, which is becoming physically important experimentally.
\end{abstract}

\newpage

\section{Introduction}

A two-level system (2LS) interacting with a quantum field -- electromagnetic
field (EMF) in particular -- has proven to be a very useful model for a wide
range of problems from  atomic-optical  \cite{WM,MW,Scu,Wei,VW,CPP,Car}
and condensed matter \cite{Leg87,Weiss}  processes to quantum computation \cite{QComp}.
For the latter application stringent limits in maintaining the coherence of the
the 2LS (called qubits)  are required . This
prompted us  to revisit the theoretical structure of the 2LS  model,
paying special attention to its coherence properties. Treatment of
spontaneous emission and relaxation are standard textbook topics,
whereas decoherence and dissipation, especially in the context of
quantum computation, are the focus of more recent investigations
\cite{Unr95,PSE,ZanRas,PleKni,SchMil,VioLlo,Gar,Zan}.

Because of the familiarity of the model (see II.A) and its theoretical and practical
values, we do not need to emphasize the general motivation, but can go right
to the point about the aim and results of this paper.
The description of this system generally comprises of  two parts:
1) Spontaneous emission in the 2LS, and
2) Decoherence due to the interaction of 2LS with the EM field, treated as
a bath.
The first part allows little room for disagreement, as it can be obtained
from elementary calculations. The second part on decoherence is more subtle.

Environment-induced decoherence \cite{envdecrev} has been studied extensively
in recent years primarily based on models of quantum Brownian motion (QBM)
\cite{FeyVer,CalLeg,GSI,UnrZur,HPZ,PHZ,HM94}
for the interaction of a simple harmonic oscillator (Brownian particle)
with a harmonic oscillator bath (HOB) at a finite temperature,
leading to a reasonably good understanding of its characteristic features.
Decoherence of a 2LS in an EM field has been studied by a number of authors, notably
\cite{Unr95,PSE,VioLlo}, and their dissipative and decoherent behavior
are reported to be similar to that of a QBM in a harmonic oscillator bath.
The progression in three stages --  quiescent, vacuum fluctuation-dominated
and thermal fluctuation-dominated, separated  by the cutoff frequency
and the thermal de Broglie frequency (wavelength)
 -- are indeed characteristic of the QBM results
\cite{CalLeg,UnrZur,HPZ,PHZ,HZ,AH}.

Our findings, in contrast,  are in stark
disagreement from that reported in the literature. We work with the standard
2LS-EMF model \cite{WM} and obtain an exact master equation for depicting
non-Markovian dynamics. Solution of this equation for the reduced density matrix
of the 2LS shows that the decoherence rate is close to the relaxation rate.  
This is in first appearance rather couter-intuitive,
and different from all previous findings. 
Upon careful deliberation we realize that the `intuition' researchers (including
us at the start) have acquired for dissipation and decoherence are based on the 
QBM model which influenced  the choice of model in the investigation of decoherence 
for a 2LS.  However, we find that such a commonly
invoked intuition for QBM in a HOB fails to apply
to that of a two-level atom (2LA) interacting with an electromagnetic field (EMF)
with the commonly assumed type of resonance coupling in quantum optics.\\

\noindent {\it Decoherence in QBM}\\

Physically, when we say that decoherence of the system of a Brownian 
oscillator proceeds in a very short time as  it is brought in  contact  with 
an environment, a HOB at some temperture, we are usually conjuring a 
model with bilinear \cite{FeyVer}
(or polynomial \cite{HPZ}) coupling of the oscillator-bath coordinates,
and a ohmic or subohmic spectral function \cite{CalLeg} in the bath.
Intuitively, the bath needs to have many degrees of freedom, preferably
acting independently of each other so incooperatively that the phase information 
in the system will be dispersed to the largest extent amongst the many bath
degrees of freedom and affords little chance or takes inordinately long time to be
revived or reconstituted (recoherence \cite{recoh}).  
The opposite picture (of very long decoherence time)
is exemplified by  two coupled subsystems  where no coarse-graining is
introduced, or for system-environment couplings which  maintain some high level of
coherence, or for an environment whose degrees of freedom  have  long correlation times 
like in a zero temperature, supraohmic bath. The case of a (spin) particle or
(plasma) wave   interacting with an averaged (collective) variable from the environment,
 such as the mean field,  showing  Landau damping in Vlasov dynamics is another example 
\cite{HuPav,HPM}. Just as in  the spin echo phenomena (e.g,  Chap. 3
\cite{Wei}),  the basic physics in this  case is not  dissipation in  the  Boltzmann 
sense,  but  statistical mixing \cite{Ma}.  We will see that this  example is
of more physical  relevance to our problem than the QBM.\\

\noindent {\it Coherence in the 2LS}\\

For the 2LA-EMF system, one clear distinction between an EM field as an
environment and a  system of harmonic oscillators as  bath is that the field 
(coupled to a detector) 
has an intrinsic spectral density function, not to be chosen arbitrarily. 
For example, it has been shown \cite{HM94}
that a conformal scalar field in two dimensions coupled to a monopole detector
has an  Ohmic character while in four dimensions it is supraohmic . 
 Barone and Caldeira \cite{CalBar} showed  that the spectral density function
for EM fields with momentum coupling to an oscillator detector  is supraohmic.
These density functions  would show very different
decoherence  behavior from the high temperature  Ohmic HOB  used in
many discussions of decoherence, the latter case is what the general folklore
is based on. But the most important  distinction  from QBM is that the 2LA 
couples with the EMF in the discrete number basis for the field, unlike the 
continuous amplitude basis in the QBM. 
This fact (which is true in the rotating wave and dipole
approximation) implies that the 2LS plus EMF system is a {\it resonant} one.
Hence even though the EM field has just as many (large number of)  modes 
as the HOB,  only a very small fraction 
of them in a narrow  range of the resonance frequency are efficiently coupled
to the atom. This is the root cause for the very different qualitative behaviour 
between the QBM and the 2LS as far as decoherence  is concerned.

One extreme case is that of a single mode field described by the
Jaynes-Cummings model, where Rabi nutation takes place and 
the atom-field  remains largely a coherent system.
(For a coherent field, the probability for the atom to be found in the
excited state at time t regardless of the state of the field) obeys a  Poisson distribution. 
This distribution in the photon number induces a  spread in the Rabi frequencies, 
and causes collapse and revival of the  Rabi nutation. These are distinct  features of quantum
coherence \cite{WM}.)  Adding all modes to the field we see  spontaneous emission 
and the decay of the atom.
The probability of an initially excited atom (remaining in the excited state)
decays exponentially in the Wigner-Weisskopf form  (characteristic of
Markovian processes) with relaxation time constant  $\Gamma$. For purely radiative 
decay the decay time $T_1$ of the inversion is half  the decay time $T_2$ of the 
polarization. There is no
large order of magnitude differences between dissipation and decoherence time
(which in  typical QBM high temperature conditions could be as high as 40  \cite{envdecrev}).
In fact it is perhaps inappropriate to talk about dissipation for a 2LA-EMF
system because the conditions for a bath to actuate such a process is lacking. The
transition from excited to ground state is closer in nature to relaxation  (in the spin echo
sense) than dissipation. In a cavity where excitation of the atom from the field (absorption)
balances  with  emission, it is more appropriate to refer to the  resonant state 
of the atom-field as a coherent system.
In these senarios the distinction between QBM and 2LA cannot be clearer. \\

\noindent {\it Difference between QBM and 2LS}\\

So what led earlier authors to make the qualitative claim that  2LS decoheres easily?
We think the confusion  arises when the picture of  QBM dissipation and
decoherence is grafted on the 2LA-EMF system indiscriminantly.
If the field which acts as the environment is a phonon field (from ion vibrations, 
see, e.g., \cite{Gar}) and if the coupling is of the non-resonant type, 
then there is no disagreement.  Decoherence should follow the QBM pattern
as reported by many authors \footnote{Even in such cases, one also needs to 
pay closer attention to the QBM behavior
than what has been accorded for this model.
Subtle points unnoticed before include, e.g., the
imposition of a high frequency cutoff and Ohmic spectral function
which restricts to a Markovian behavior \cite{CalLeg} can  lead to a violation of the
positivity of the reduced density matrix \cite{HPZ},
the violation of the fluctuation-dissipation relation \cite{UnrZur},
and the prolongation of coherence in a low temperature supraohmic
bath \cite{HPZ}. They deserve more attention in the theoretical design of
cavity qubit computers.}. 
Such sources (including atomic collisions in a cavity \cite{Car})
can be important for some setups. However,  when
one claims that the EM field can decohere a 2LS (with which it is coupled
in a resonant way, as in the standard model) that is where we disagree.

Quantitatively, the model for the 2LS used by most authors for the discussion
of decoherence  inspired by QBM type of behavior
has the atom in a $\sigma_z$  state (the diagonal Pauli matrix) coupled to the
field mode operators $\hat b^\dagger, \hat b$.  This type of coupling term (call it 
$\sigma_z$ type for convenience)   commutes with the
Hamiltonian of the system, and  admits a diagonalization 
in the eigenbasis of the Hamiltonian. The field is coupled to the 
atom as a whole and thus is  insensitive to the 2 level transition activity.
In particular it  does not probe the resonance or coherent properties
of the two level atom, which is the most important feature,
for quantum computation. 
By contrast the standard model for 2LA-EMF which we studied 
has a $\sigma_\pm$ coupling (call it standard coupling) 
to the field modes which highlights the
2 level activity of the atom and the field. This  coupling considered in 
the standard model is indispensible,  i.e., it {\it cannot}  be removed 
from the two-level atom as it {\it defines} it 
and will be present in any realistic situation.
What then is the origin of the QBM type of contribution to the 2LS ?

If one accepts an environment other  than the EM field, the question 
comes down to the characteristics of the experimental apparatus. For
well- prepared ion traps  we would expect it to be rather unimportant.
If  the EM field is the only field present,  we can still  ask if 
a QBM type of coupling term with the EM field would appear, 
and if yes,  how strong would its effect be?  This would be a
useful way to accomodate the two different types of  coupling terms.
\par
Recall that the standard model is derived  under the dipole and rotating wave 
approximation. 
 In the next section we will show that the
$\sigma_z$ type of coupling appears only in the next order expansion 
after the dipole approximation. Since these are good approximations 
for a large class of atomic states when the atom is nonrelativistic,
the contribution from the QBM type of coupling used in  \cite{Unr95,PSE,VioLlo}
should be  negligible and its ensuing decoherent effect insignificant.
In this sense the EM field does not in leading order of approximation act like a bath 
in the QBM way,  and coherence  in a 2LA-EMF system  is quite well preserved 
(excepting other processes, e.g. \cite{SchMil,PleKni}).  

Our puzzle  over the result on decoherence in the 2LS  reported in the literature
was what prompted us to begin this study. Without letting any familiar and 
convenient analogy influence our judgement, and without any preconceived
notion,  we  choose to perform a first-principles calculation
of the two-level atom (2LA)-electromagnetic field (EMF) system
making as few assumptions  and covering as wide a range of
conditions as possible. 
 We use the influence functional method \cite{FeyVer} to take into account
the full backreaction of the field on the atom, while adopting 
Grassmannian variables for the 2LA and the coherent state representation for
 the EMF. We find exact master equations for the full
(non-Markovian) dynamics in the cases of a free quantum field and a cavity field
at zero temperature. 

In the next section we present the model and the formalism.  A 
detailed derivation of our model is contained in Appendix A.
In Sec. 3 we derive the master equations. In Sec. 4 we study
different mode composition of the field, including that of an
atom in a cavity.  We end in Sec. 5 with a discussion
of our findings and their implications.

This is the first in a series of papers on 2LAtom and quantum decoherence.
The  subsequent papers will treat 2LA- EMF interaction at finite temperature,
for EM fields in a coherent and squeezed state, 
and for multipolar models (where coupling other than the minimal is assumed).
We will also tend to collective qubit systems and moving atoms interacting with an EM field.
These results will have corresponding applications in atom optics and
quantum computation problems.

\section{The Influence Functional}

\subsection{The Model}

Our model for atom-field interaction is the standard one (see Appendix A for
details)  \cite{WM,MW,Wei}
\footnote{Our Hamiltonian is given in the so-called
minimal  coupling (MC) as different from the multipolar coupling (MP)
\cite{CPP}, which may be more relevant to atoms in a cavity
because the expicit Coulomb interaction between the atom and its
image charge is removed.}.
The total Hamiltonian for a (stationary) atom interacting with a
quantum electromagnetic field (EMF) under the dipole, rotating wave (RW)
and two-level (2L) approximation is given by

\be
{\hat H} = \hbar \o_0 {\hat S}_z
+ \hbar \sum_\bk \left[ \o_\bk {\hat b}_\bk^{\dagger} {\hat b}_\bk
+    \left(g_\bk  S_+ {\hat b}_\bk  +\bar g_\bk  S_- {\hat  b}_\bk^{\dagger} \right) \right]
\ee
where
${\hat b}_\bk^{\dagger},{\hat b}_\bk$ are the creation and annihilation
operators for the kth normal mode with frequency $\o_\bk$
of the electromagnetic field (thus for the field vacuum ${\hat b}_\bk|0\rangle
= 0, [{\hat b}_\bk, {\hat b}_{\bk'}^{\dagger}] = \delta_{\bk,\bk'}$,
 for all $\bk$.), and $ \o_0= \o_{21}$ is the frequency between the two levels.
Here 
$$
\hat S_z = \ha \hat \s_z,
\hat S_\pm = \hat \s_\pm  \equiv \ha (\hat \s_x \pm i \hat \s_y)
$$
where $\s_{x,y,z}$ are the standard 2x2 Pauli matrices with $\s_z = diag (1, -1)$,
etc.
The coupling constant $g_\bk \equiv  d_{21\bk} f_\bk (\bX)$  where
\begin{equation}
d_{ij\bk} \equiv -\frac{i\o_{ij}}{\sqrt{2\hbar\o_\bk\epsilon_0 V}}\bd_{ij}
\cdot \hat {\bf e}_{\bk \s}
\end{equation}
and  $\bd_{ij} \equiv e \int {\bar \phi_i}{\bf x}\phi_j d^3x$ is the dipole
matrix element between the eigenfunctions $\phi_i$ of the electron-field
system,
 $\hat {\bf e}_{\bk \s}$ is the unit polarization vector ( $\s =1, 2$ are the
two polarizations), and $f _\bk (\bx)$ is the spatial mode functions of
the vector potential of the electromagnetic field (in free space,
$f_\bk (\bx) = e^{-i \bk \cdot \bx}$,  $V$ is the volume of
space.). Under the dipole approximation $f_\bk$ 
is evaluated at the position of the atom $\bX$.
Since $ \bd_{ij} = {\bar \bd}_{ji}$, ${\bar d}_{ij\bk} = d_{ji\bk}$, we will
choose a mode function representation such that $g_\bk$ is real.

To see how this could possibly be related to the $\sigma_z$ type of coupling
with Hamiltonian (used by e.g., \cite{PSE,VioLlo} for the study of decoherence in 2LS)
\be
{\hat H} = \hbar \o_0 {\hat S}_z
+ \hbar \sum_\bk \left[ \o_\bk {\hat b}_\bk^{\dagger} {\hat b}_\bk
+ \hbar \sigma_z   \left(\bar  g_\bk {\hat b}_\bk  + g_\bk  {\hat  b}_\bk^{\dagger} \right) \right]
\ee
we examine  the next term after the dipole approximation in (A.15).
This has a  contribution to $g_{ij \bk}$ even when $i=j$ which is  equal to 
$$
g_{ii {\bf k}} = c_{{\bf k}} {\bf k} \cdot {\bf q}_i
$$
where
$$
{\bf q}_i = \sum_{\sigma} \int \bar{\phi}_i {\bf \delta x} ({\bf p} \cdot {\bf
\hat{e}}_{{\bf k} \sigma} ) \phi_i dx^3
$$
and $c_k$ is a constant given by   
$$
c_{\bf k} = - \frac{e}{m} (2 \hbar  \omega_{{\bf k}} \epsilon_0 V)^{-1/2}
$$
This generates  an additional coupling term
$$
       \sum_{\bf k} \sigma_z ( g_{1{\bf k}} b_{{\bf k}} +\bar  g_{1 {\bf k}}
b^{\dagger}_{{\bf k}}) + 1 (g_{2{\bf k}} b_{{\bf k}} +\bar  g_{2 {\bf k}}
b^{\dagger}_{{\bf k}})
$$
where
$$
g_{1 {\bf k}} = g_{11{\bf k}} - g_{22 {\bf k}}, \,\,\,
g_{2 {\bf k}} = g_{11{\bf k}}+ g_{22 {\bf k}}
$$
This  gives  the lowest order  $\sigma_z$ type of
coupling in a 2LA -EMF system.
The ratio of  the coupling $g_{1{\bf k}}$ of the $\sigma_z$ type in Eq.  (II.3)to 
the dipole coupling $g_\bk$  in  Eq. (II.1) is
\begin{equation}
|g_{1 {\bf k}}/g_{ {\bf k}}| = | \frac{{\bf k} ({\bf q}_1 - {\bf q}_2)}{ m
\omega_\bk  d_{12}}| \leq \frac{\omega_\bk |{\bf q}_1 -{\bf q}_2|}{m \omega_\bk d_{12}}
\end{equation}
Thus the $\sigma_z$ type of coupling generated from the 2LA- EMF interaction 
will be  significant only for very  high frequencies $\omega_k$ of
the EM field, a  point intuitively clear from the meaning of the dipole approximation.
\\ 
\subsection{Grassmannian Variables and Coherent State Integrals}

Since Feynman and Vernon \cite{FeyVer} invented the 
influence functional method this formalism has been applied  to
treat the Brownian motion of a harmonic oscillator
interacting with a harmonic oscillator bath by many authors
\cite{CalLeg,Weiss,GSI,HPZ}. 
The two - level system in tunneling has been discussed in detail by Leggett et
al \cite{Leg87},
but the derivation of a master equation by this method which can
traverse the non Markovian regimes has not yet been carried out.
We shall perform such a calculation for a two level system, with the aid of
Grassmaninn variables convenient for treating fermions, and
the coherent state representation in a path integral form.
We construct the coherent state of the combined atom-field system as
\begin{equation}
|\{z\},\eta  \rangle =|\{z\}\rangle \times |\eta  \rangle
\end{equation}
where  $|z\rangle $, $z$ a complex number, denotes the EM field coherent
states
and $|\eta$, $\eta$ a Grassmannian or anticommuting number, denotes the
electron coherent state.  The transition amplitude between the initial
state (i) at $t=0$ and the final state (f) at $t=t_f$ is
expressed formally as \cite{NegOrl} (here we suppress the index $k$,
\begin{equation}
\langle \bar{\eta }_{f},\bar{z}_{f};t|\eta  _{i},z_{i};0\rangle =\int
DzD\bar{z%
}D\eta  D\bar{\eta }e^{\frac{i}{\hbar}S[z,\bar{z},\eta  ,\bar{\eta }]}
\end{equation}
where the action is 
\begin{equation}
\frac{i}{\hbar}S[z,\bar{z},\eta  ,\bar{\eta }]=\bar{z}z(t)+\bar{\eta }\eta
(t)-\int_{0}^{t}ds(%
\bar{z}\dot{z}+\bar{\eta }\dot{\eta }+\frac{i}{\hbar}
\hspace{0.1cm}H(\bar{\eta },\eta  ,%
\bar{z},z)
\end{equation}
Here $H$ is the Q-symbol of the Hamiltonian \cite{Kib}
and there is an implied summation over field modes.
\begin{equation}
H(\bar{\eta },\eta  ,\bar{z},z)=\hbar \left( \sum_k  \omega _k  %
\bar{z}_k  z_k  -g _k  (\bar{z}_k  \eta  +\bar{%
\eta }z_k)  +\omega_0 \bar{\eta }\eta  \right)
\end{equation}
In  (II.8) we have substracted a constant term $\frac{1}{2}\omega_0 1$ to (II.1) 
so that the ground state now has zero energy.  Henceforth we set $\hbar=1$.

The Hamiltonian in equation (2.8) is not a c-number function; it has terms 
that are odd. One might then question the validity of equation (2.6) for the 
path integral; it clearly exists as a formal expression, but its evaluation 
with a saddle point method, that is based on the Hamiltonian of equation (2.8) 
might be problematic. 
We dispell this doubt for the vacuum case with an operaytor method proof of the master equation in Appendix B. It shows that at least for the vacuum case tha saddle point evaluation yields the correct result. The general cases need separate considerations.
For  many qubits coupled to the EM field vacuum, we believe that the path integral method yields a simpler treatment than the operator method.

The integration is over all paths satisfying
\begin{eqnarray}
z(0)&=&z_{i} \hspace{3cm}\bar{z}(t)=\bar{z}_{f} \\
\eta  (0) &=&\eta  _{i} \hspace{3cm} \bar{\eta }(t)= \bar{\eta }_{f}
\end{eqnarray}
We assume initially that the density matrix of the total system+environment
is factorizable $ \hat \rho (0) = \hat \rho_e (0) \otimes \hat \rho_b (0)$.
Only at that time would $z$ and $\eta$ be pure complex and Grassmannian
numbers respectively. As the system evolves, both $\eta  $ and $z$
contain Grassmann and c-number parts. The mixing of even and odd parts
(note $g_k$ is odd) comes about as the initially factorized atom state
becomes "dressed".

In the open system philosophy, as we are interested in the averaged effect
of the field on the atom, the atom is considered as the `system' while
the field as the `environment'.
The path integral is performed over the variables $z$, while
$\eta, \bar{\eta }$ are treated as external sources.
When only one field mode is considered, we have
\be
\langle \bar{z}_{f};t | z_{i};0\rangle_{\eta  ,\bar{\eta }}
= \int Dz D\bar{z}%
\exp \left\{ \bar{z}z(t)-\int_{0}^{t}ds(\bar{z}\dot{z}+i(\omega \bar{z}z
-g (\bar{z}\eta  +\bar{\eta }z)(s)\right\}
\ee
with summation over paths satisfying the boundary condition (II.9 ) for $z$.
We use the saddle point method. Minimizing the action
yields the following equations
\begin{eqnarray}
\dot{z}+i\omega z &=& -ig \eta  \\
\dot{\bar z}-i\omega {\bar z} &=& ig \bar{\eta }
\end{eqnarray}
with solutions
\begin{eqnarray}
z(s) &=& z_{i}e^{-i\omega s}-ig \int_{0}^{s}ds^{\prime }e^{-i\omega
|s-s^{\prime}|}
\eta  (s^{\prime }) \\
\bar{z}(s) &=&\bar{z}_{f}e^{-i\omega (t-s)}+ig \int_{s}^{t}ds^{\prime}
e^{-i\omega |s-s^{\prime }|}\bar{\eta }(s^{\prime })
\end{eqnarray}
Using these for the transition amplitude (II.11 ) with the minimum value for the action,
we obtain
\begin{eqnarray}
\langle \bar{z}_{f};t | z_{i};0\rangle_{\eta  ,\bar{\eta }}
 = \exp \left\{ \bar{z}_{f}z_{i}e^{-i\omega t}
-ig \left[ \bar{z}_{f} \int_{0}^{t}ds e^{-i\omega (t-s)}\eta  (s)
\right.\right.  \nonumber \\
   \left. \left.       +\int_{0}^{s}ds e^{-i\omega s}\bar{\eta }(s)z_{i}
 \right]
 -g ^{2}\int_{0}^{t}ds\int_{0}^{s}ds^{\prime }e^{-i\omega
|s-s^{\prime }|}\bar{\eta }(s)\eta  (s^{\prime }) \right\}
\end{eqnarray}
A prefactor in the coherent state path integral is 
equal to one. Now the influence functional due to  this single mode reads
\begin{eqnarray}
{\cal F}[\eta  ,\bar{\eta };\eta^{\prime }\bar{\eta }^{\prime }] &=&\int
\frac{d\bar{z%
}_{i}dz_{i}}{\pi }\frac{d\bar{z^{\prime }}_{i}dz_{i}^{\prime }}{\pi }\frac{d%
\bar{z}_{f}dz_{f}}{\pi } e^{-\bar{z}_i z_i - \bar{z}_i' z_i' - \bar{z}_f z_f} \nonumber \\
&&\times \langle \bar{z}_{f};t|z_{i};0\rangle _{\eta  ,\bar{\eta }}\hspace{%
0.2cm}\langle \bar{z}_{i}|\rho _{0}|z_{i}^{\prime }\rangle \hspace{0.2cm}%
\langle \bar{z}_{i}^{\prime };0|z_{f};t\rangle _{\eta  ,\bar{\eta }}
\end{eqnarray}
where the completeness relation for (unnormalized) coherent states has been
used
\begin{equation}
\int \frac{d\bar{z}_{i}dz_{i}}{\pi }e^{- \bar{z}z} |z\rangle \langle
\bar{z}|=1
\end{equation}
Writing with an obvious identification 
\begin{equation}
\langle \bar{z}_{f};t|z_{i};0\rangle _{\eta  ,\bar{\eta }}=\exp \left(
A\bar{z}%
_{f}z_{i}+i\bar{z}_{f}\beta +i\bar{\gamma}z_{i}+D\right)
\end{equation}
we can use the identity 
\begin{equation}
\int \frac{d\bar{z}_{i}dz_{i}}{\pi
}e^{-\bar{z}z+\bar{f}z+\bar{z}f}=e^{\bar{f}f}
\end{equation}
 to obtain 
\begin{equation}
{\cal F}[\bar{\eta },\eta  ,\bar{\eta }^{\prime },\eta  ^{\prime
}]=e^{\bar{\beta}^{\prime }
\beta -(D+D^{\prime })}
\end{equation}
for an initial vacuum state $\hat{\rho}_{0}=|0\rangle \langle 0|$.
Substituting, we get the contribution to the influence functional from one
mode 
\begin{eqnarray}
{\cal F}_k  [\bar{\eta },\eta  ,\bar{\eta }^{\prime },\eta  ^{\prime }]
&=&\exp \left\{ g_k^{2}\int_{0}^{t}ds\int_{0}^{s}ds^{\prime }
\left[ \bar{\eta}^{\prime }(s)\eta  (s^{\prime })e^{-i\omega _k
(s-s^{\prime})}\right.\right.
\nonumber \\
& & \left.\left.+\bar{\eta}^{\prime }(s^{\prime })\eta
(s)e^{i\omega_{k}(s-s^{\prime })}
-\bar{\eta }(s)\eta  (s^{\prime })e^{-i \omega_{k}(s-s^{\prime })}-
\bar{\eta }^{\prime }(s^{\prime })\eta  (s)e^{i\omega_k
 (s-s^{\prime })}\right] \right\}
\end{eqnarray}
The influence functional for all modes
${\cal F}=\prod_{k}{\cal F}_k  $ is finally given by
\begin{eqnarray}
{\cal F}[\bar{\eta },\eta  ,\bar{\eta }^{\prime },\eta ^{\prime }] &=&\exp
\left\{
\int_{0}^{t}ds\int_{0}^{s}ds^{\prime }\left(\mu  (s-s^{\prime })[\bar{\eta }%
^{\prime }(s)-\bar{\eta }(s)]\eta  (s^{\prime }) \right.\right.
 \nonumber \\
&&  \left.\left.+\mu ^{*}(s-s^{\prime })\bar{\eta}^{\prime }(s^{\prime })
[\eta(s)-\eta ^{\prime}(s)]\right)\right\}
\end{eqnarray}
in terms of the kernel
\begin{equation}
\mu  (s)=\sum_k  g_k^{2}e^{-i\omega _k  s}
\end{equation}

\section{The master equation}

\subsection{The reduced density matrix propagator}

Having computed the influence functional we have an expression for the
reduced density matrix propagator 
\begin{eqnarray}
J(\bar{\eta }_{f}\eta  _{f}^{\prime };t |\bar{\eta }_{i}^{\prime }\eta
_{i};0)
=\int D\bar{\eta }D\eta  D\bar{\eta  ^{\prime }}D\eta  ^{\prime }\exp
\left\{
\bar{\eta }\eta  (t)+\bar{\eta }^{\prime }\eta  ^{\prime }(t)
-  \int_{0}^{t}ds ( \bar{\eta }\dot{\eta }+\bar{\eta }^{\prime } \dot \eta
^{\prime }+i\omega \bar{\eta }\eta  -i\omega \bar{\eta }^{\prime
}\eta^{\prime} (s)
 \right.  \nonumber \\
 \left. + \int_{0}^{t}ds\int_{0}^{s}ds^{\prime } \left( \mu  (s-s^{\prime
})[%
\bar{\eta }^{\prime }(s)+\bar{\eta }(s)]\eta  (s^{\prime })
+\mu  ^{*}(s-s^{\prime })\bar{\eta }^{\prime }(s^{\prime
})[\eta  (s)+\eta  ^{\prime }(s)]\right) \right\}
\end{eqnarray}
where the summation over all paths obey the boundary conditions (II.10) and
\be
\bar{\eta }^{\prime }(0)=\bar{\eta }_{i}^{\prime }
\hspace{3cm} \eta^{\prime }(t)=\eta_{f}^{\prime }
\ee
We can compute the path integral with saddle point evaluation and get
\begin{eqnarray}
\dot{\eta }+i\omega \eta  +\int_{0}^{s}ds^{\prime }\mu  (s-s^{\prime })\eta
(s^{\prime }) &=&0\hspace{4cm} \\
\dot{\bar{\eta }^{\prime }}-i\omega \bar{\eta }^{\prime
}+\int_{0}^{s}ds^{\prime }\mu  ^{*}(s-s^{\prime })\bar{\eta }^{\prime
}(s^{\prime }) &=&0\hspace{4cm} \\
\dot{\eta }^{\prime }+i\omega \eta  ^{\prime }+\int_{0}^{s}ds^{\prime }\mu
(s-s^{\prime })\eta(s^{\prime })-\int_{s}^{t}ds^{\prime}\mu^{*}(s-s^{\prime})
[\eta (s^{\prime })+\eta^{\prime }(s^{\prime })] &=&0 \\
\dot{\bar{\eta}}-i\omega \bar{\eta }+\int_{0}^{s}ds^{\prime
}\mu^{*}(s-s^{\prime})
\bar{\eta }^{\prime }(s^{\prime })-\int_{s}^{t}ds^{\prime }\mu
(s-s^{\prime })[\bar{\eta }^{\prime }(s^{\prime })+\bar{\eta }(s^{\prime
})]&=&0
\end{eqnarray}
It will turn out that only the solution of the first two of these equations
will contribute to the path integral. We will therefore write 
\be
\eta  (s)=\eta  _{i}u(s)\hspace{3cm}\bar{\eta }^{\prime }(s)=\bar{\eta }%
_{f}^{\prime }\bar{u}(s) 
\ee
where $u$,$\bar{u}$ are the solutions to equations (III.3),(III.4) under the
condition 
\be
u(0)=\bar{u}(t)=1
\ee
Now equation (III.3) is a linear integrodifferential equation of first order
and as such can be solved with the use of the Laplace transform and the
convolution theorem. It is easy to show that 
\begin{equation}
u(s)={\cal L}^{-1}\left( \frac{1}{z+i\omega +\tilde{\mu }(z)}\right)
=\frac{1%
}{2\pi i}\int_{c-i\infty }^{c+i\infty }\frac{dze^{zs}}{z+i\omega +\tilde{\mu
}(z)}
\end{equation}
where $\tilde{\mu }(z)$ is the Laplace transform of the kernel (II.24) and $c$
is
a real constant larger than the real part of the poles of the integrand. It
turns out that this function $u(s)$ contains all necessary information for
the computation of the density matrix propagator.
Substituting our expressions (III.8) and (III.9)) in (III.1) we can obtain
the following expression for the propagator 
\begin{equation}
J(\bar{\eta }_f \eta ^{\prime}_f;t| \bar{\eta }^{\prime}_i \eta _i;0) = \exp
\left( \bar{\eta }_f \eta _i u(t) + \bar{\eta} ^{\prime}_i
\eta ^{\prime}_f \bar{u}(t) - [1 -\bar{u}(t) u(t)] \bar{\eta }^{\prime}_i
\eta _i \right)
\end{equation}
Since we are using coherent state path integrals we have departed in our
evaluation from the standard saddle point approximation used on
configuration space path integrals. In these cases , the standard procedure 
is to distinguish the imaginary part of the kernel as corresponding to 
dissipation and consider only its contribution when performing the 
saddle point evaluation. The resulting equations are then the classical 
dissipative equations of motion. But in the case of the coherent state 
path integral, there is no correspondence between extrema of the action 
and actual classical paths.  Hence, there is no sense in splitting the kernel 
$\eta$ into real and imaginary part, and the saddle point evaluation 
should be carried out for the whole of the exponential.

\subsection{Master equation for a field in a vacuum state}

It is a standard procedure now to find the master equation \cite{CalLeg,HPZ}.
We compute the time derivative of the propagator: 
\begin{equation}
\dot{J} = (\dot{u} \bar{\eta }_f \eta _i + \dot{\bar{u}} \bar{\eta }%
_i^{\prime}\eta _f + \frac{d (\bar{u}u)}{dt} \bar{\eta }^{\prime}_i \eta _i)
J
\end{equation}
The next step is to remove from the above equation the dependence on the
initial values. This is done with the use of the following identities 
\begin{equation}
\eta _i J = \frac{1}{u} \frac{\delta J}{\delta \bar{\eta }_f} \hspace{3cm}
\bar{\eta }^{\prime}_i J = \frac{1}{\bar{u}} \frac{\delta J}{ \delta
\eta _f^{\prime}}
\end{equation}
Note that we are suppressing (for ease of notation) symbols denoting left
or right Grassmann differentiation. In all our expressions we implicitly
assume that differentiation with respect to $\eta $ is always right and with
respect to $\bar{\eta }$ always left.

For the density matrix at time $t$ 
\begin{equation}
\rho_t(\bar{\eta }_f,\eta ^{\prime}_f) = \int d \bar{\eta }_i
d
\eta _i e^{-\bar{\eta}_i \eta_i}
d \bar{\eta }^{\prime}_i
d
\eta _i^{\prime} e^{-\bar{\eta}'_i \eta'_i}
J(\bar{\eta }_f \eta ^{\prime}_f;t| \bar{\eta }^{\prime}_i
\eta _i;0) \rho_0(\bar{\eta }_i,\eta^{\prime}_i)
\end{equation}
we obtain the evolution equation 
\begin{equation}
\frac{\partial}{\partial t} \rho = \frac{\dot{u}}{u} \bar{\eta } \frac{\delta
\rho}{\delta \bar{\eta }} + \frac{\dot{\bar{u}}}{\bar{u}} \frac{\delta
\rho}{%
\delta \eta } \eta  + \frac{\frac{d}{dt}(\bar{u}u)}{\bar{u}u} \frac{\delta^2
\rho}{\delta \eta  \delta \bar{\eta }}
\end{equation}
This is one of our main results: The master equation for the two-level atom
interacting with a an environment of electromagnetic field at its vacuum
state. The effect of the field is contained within the function $u$ which
can be determined by the solution of equation (III.3) or equivalently by the
computation of the contour integral (III.9). In the next section we are going
to find explicit expressions for $u$ for particular choices of the field
configuration.

Let us return for the moment to equation (III.14) and write this in an
operator language. It is easy to verify that: 
\begin{equation}
\bar{\eta } \frac{\delta \rho}{\delta \bar{\eta }} = \Sigma_+ \rho
\hspace{2cm}
\frac{\delta \rho}{\delta \eta } \eta  = \rho \Sigma_+ \hspace{2cm} \frac{%
\delta^2 \rho}{\delta \eta  \delta \bar{\eta }} = S_+\rho S_-
\end{equation}
where $\Sigma_+ = (1 + \sigma_z)/2$.
If we write 
\begin{equation}
\frac{\dot{u}(t)}{u(t)} = \Gamma (t) +i \Omega(t)
\end{equation}
the master equation reads 
\begin{equation}
\frac{\partial}{\partial t} \rho = -i  [\Omega(t) S_+ S_-,\rho] + \Gamma (t)
\{S_+S_-,\rho \} - 2 \Gamma (t) S_- \rho S_+
\end{equation}
where 
\begin{eqnarray}
H(t) = \Omega(t) \Sigma_+                        
\end{eqnarray}
The first term corresponds to the unitary Hamiltonian evolution, only now
the effect of the environment has induced a time dependent shift in the
value of the frequency, the second term is time dependent dissipation and
the third corresponds to noise.

\subsection{Spontaneous emission}

To show how the standard results are regained, and to understand the
meaning of the new function in the master equation, let us consider
the physical process of spontaneous emission.  Start with a generic
initial density matrix
\be
\rho =\left( 
\begin{array}{cl}
1-x & y \\ 
y^* & x
\end{array}
\right) 
\ee
its corresponding Q-symbol is 
\begin{equation}
\rho (\bar{\eta },\eta  )=x+y^{*}\eta  +y\bar{\eta }+(1-x)\bar{\eta }\eta
\end{equation}
If we evolve it with the density matrix propagator (III.10) we obtain for the
state at time $t$ 
\begin{equation}
\rho _{t}(\bar{\eta },\eta  )= 1 - \bar{u}u(1-x) 
+\left(
\bar{u}%
y^{*}\eta  \right) +\left( u\bar{\eta }y\right) +\left( \bar{u}u(1-x)\right) \bar{\eta} \eta
\end{equation}
corresponding to 
\be
\rho _{t}=\left(
\begin{array}{cl}
\bar{u}u(1-x) & uy \\ 
\bar{u}y^* & 1-\bar{u}u(1-x)
\end{array}
\right) 
\ee
Considering the case $x=y=0$ we get for the probability of spontaneous
emission 
\be
P(1\rightarrow 0,t)=1-\bar{u}u
\ee
Also we should remark that the rate of decoherence in the energy eigenstates
is governed by the absolute value of the function $u$ (the off- diagonal
terms).
 But on the other hand $u$ itself determines the rate of nergy flow from the 
atom to the environment. Hence for our particular choice of initial state
(vacuum) 
we find that decoherence and relaxation time are essentially identical .
 We shall use this equation to study decoherence
in an ion trap in a later paper.

\section{Field modes and analytic u(t) }

Our master equation (III.17) depends solely on the function $u(t)$ , which in
its turn is determined by the kernel $\mu (s)$ . In this section we will try
to give some analytic expressions for this function in various different
cases.

\subsection{A single mode}

To connect with known results \cite{WM}, let us start with the case when the
field
contains only a single mode with frequency $\omega _{k}=k$. Then $%
\mu  (s)$ will read
\begin{equation}
\mu  (s)=g ^{2}e^{-iks}
\end{equation}
and 
\begin{equation}
\tilde{\mu }(z)=-g \int_{0}^{\infty }e^{-sz}e^{-iks}=\frac{g ^{2}%
}{z+ik}
\end{equation}
The integrand has two poles at the solutions of the equation
\begin{equation}
z^{2}+i(\omega +k)z-\omega k+g ^{2}=0
\end{equation}
given by 
\begin{equation}
z=-i\frac{\omega +k \pm [(\omega -k)^{2}+g ^{2}]^{1/2}}{2}=-i\omega _{1,2}
\end{equation}
Hence 
\begin{equation}
u(s)=\frac{k-\omega _{1}}{\omega _{2}-\omega _{1}}e^{-i\omega _{1}s}-\frac{%
k-\omega _{2}}{\omega _{2}-\omega _{1}}e^{-i\omega _{2}s}
\end{equation}
This result is in agreement with standard ones \cite{WM}

\subsection{Infinite number of modes}

Now we consider the case of the vacuum electromagnetic field in free space
i.e. not constrained by a cavity. The kernel will read then (using equation
(II.24)) 
\begin{equation}
\mu (s) = 2 \lambda^2 \int \frac{d^3k}{(2 \pi)^3} k^{-1} e^{-iks} =
\frac{\lambda^2}{%
\pi^2} \int_0^{\infty} k dk e^{-iks} = \frac{d}{ds} \nu(s)
\end{equation}
where 
\begin{equation}
\nu(s) = \frac{i\lambda^2}{\pi^2} \int_0^{\infty} dk e^{-iks}
\end{equation}
Note the factor of $2$ in (IV.6) coming from the two photon polarisations and
that
 we in view of (II.3) we have written $g_{{\bf k}} = \lambda \omega_{{\bf
 k}}^{-1/2}$. 
\par
Since the integral (IV.7) is not convergent, we will introduce an exponential
cut-off in the higher frequency modes. The presence of the cut-off is of
physical significance since we do not expect high electromagnetic modes to
couple with our two-level atom.

Hence the kernel $\nu $ will read 
\begin{equation}
\nu (s)=\frac{i\lambda^{2}}{\pi ^{2}}\int_{0}^{\infty }dke^{-iks-k\epsilon
}=\frac{%
\lambda^{2}}{\pi ^{2}}\frac{1}{s-i\epsilon }
\end{equation}
The Laplace transform of $\nu $ is then
\begin{equation}
\tilde{\nu}(z)=\frac{\lambda^{2}}{\pi ^{2}}\int_{0}^{\infty
}ds\frac{e^{-sz}}{%
s-i\epsilon }=-\frac{\lambda^{2}}{\pi ^{2}}e^{-i\epsilon z}Ei(-i\epsilon z)
\end{equation}
where $Ei$ denotes the exponential integral function analytically continued
to the complex domain. At the limit $\epsilon \rightarrow 0$ this is
essentially 
\begin{equation}
Ei(- i\epsilon z)=\gamma +\log (-i\epsilon z)+O(\epsilon )
\end{equation}
where $\gamma $ is the Euler-Macheronni constant and the logarithm is taking
values in the primary branch. Thus $\tilde{\mu }(z)$ reads ($\nu(0)$ is here
$\nu(s =0)$ 
obtained by the integration by parts of the Laplace transform)
\begin{equation}
\tilde{\mu}(z)=-\nu (0)+z\tilde{\nu}(z)=-\frac{i\lambda^{2}}{\pi ^{2}\epsilon
}-%
\frac{\lambda^{2}}{\pi ^{2}}z e^{-i\epsilon z}Ei(-i\epsilon z)
\end{equation}
Note that the cut-off $\epsilon $ affects significantly $\mu  (z)$ only at
large values of $z$, which essentially correspond to the very short time
limit, i.e the time where the two-level atom starts ''getting acquainted''
with the photon reservoir. At larger times ($t>>\epsilon $) we do not expect
the cut-off to contribute significantly in the evolution. This is a rather
typical 
behaviour in quantum Brownian motion models, provided that the ultraviolet 
cut - off of the environment is much larger than the natural frequencies of
the system.
\par
To evaluate the integral we first have to find the poles of the denominator.
We can do that in a perturbation expansion . First let us absorb the
divergent
 $\nu(0)$ factor in a frequency renormalisation.
\begin{equation}
\tilde{\omega} = \omega -\frac{\lambda^2}{\pi^2 \epsilon}
\end{equation}
so that we need find the zeros of the function  $ z  + i \omega z - 
\frac{\lambda^2}{\pi^2} e^{-i \epsilon z } Ei (-i \epsilon z)$. 
Looking for a solution in the vicinity of $z = - i \tilde{\omega}$ we find
\begin{eqnarray}
z  = - i \tilde{\omega} - \tilde{\mu}( - i \tilde{\omega}) \nonumber \\
-  i (\tilde{\omega} - \frac{\lambda^2 \tilde{\omega}}{\pi^2} 
\log(e^{\gamma} \epsilon \tilde{\omega}))- \frac{\lambda^2
\tilde{\omega}}{\pi}
 + O(\lambda^4) := - i \Omega - \Gamma
\end{eqnarray}
We have a pole with a negative real part and we can verify numerically 
(also physically expected) that there is no pole with greater real part. 
This means when evaluating the inverse Laplace transform we can ignore 
the contribution of the branch - cut at $z = 0$ (being on the right of the
pole) 
and hence after some time, where all possible other poles with absolutely 
larger value of their real part will have stopped contributing the solution
will be
\begin{equation}
u(s) = e^ {-i \Omega s - \Gamma s}
\end{equation}
This implies a Markovian time evolution and the identification of decoherence
-
 relaxation time with $ \Gamma^{-1} = \frac{\pi}{\lambda^2 \omega}$.
This is a general feature of the presence of a continuum of modes as can be
seen
 from the case of an atom transparent to all modes but the ones in a  strip ,
say $[\omega_1, \omega_2]$ containing the resonance frequency.
\par
It is easy to verify that in this case we have again
 \begin{equation}
\tilde{\mu}(z) = - i \frac{\lambda^2}{\pi^2} (\omega_2  - \omega_1) +
 \frac{\lambda^2}{\pi^2} \log \left( \frac{\omega_2 - i z}{\omega_1 - i z}
 \right)
\end{equation}
Hence defining again
\begin{equation}
\tilde{\omega} = \omega - \frac{\lambda^2}{\pi^2} (\omega_2  - \omega_1)
\end{equation}
we can find the pole at
\begin{equation}
z = - i (\tilde{\omega} - \frac{\lambda^2 \omega}{\pi^2} 
\log \left(\frac{\omega_2 - \tilde{\omega}}{\tilde{\omega} - \omega-1}
\right) ) 
- \frac{\lambda^2 \tilde{\omega}}{\pi} + O(\lambda^4)
\end{equation}

Note  that the real part of the pole comes from the presence of a minus sign 
in a logarithm of some real valued object. Hence in the case where the atom
's 
frequency is outside the strip of interacting modes there will be no
dissipation.
This feature separates us from the QBM case, characterising the atom - field
system 
as primarily a resonant one.

\subsection{Atom in a cavity}

Let us now consider the case of the atom lying within a cavity consisting of
two parallel plates at distance $L$. The field satisfies Dirichlet boundary
conditions on the surface of the plates, hence the modes in the normal
direction to the plates are multiples of $\pi /L$. The kernel then reads 
\begin{eqnarray}
\mu  (s) &=&\frac{\lambda^{2}}{2\pi L}\sum_{n}\int_{0}^{\infty }\frac{kdk}{%
(k^{2}+(n\pi /L)^{2})^{1/2}}e^{-i(k^{2}+(n\pi /L)^{2})^{1/2}(s-i\epsilon )} 
\nonumber \\
&=&\frac{1}{2\pi L}\sum_{n}\int_{|n\pi /L|}^{\infty }dke^{-ik(s-i\epsilon )}
\nonumber \\
&=&\frac{1}{2\pi L}\frac{1}{\epsilon +is}\sum_{n}e^{-i|n\pi /L|(s-i\epsilon
)} \\
&=&\frac{1}{2\pi L}\frac{1}{\epsilon +is}\frac{1+e^{-i\pi /L(s-i\epsilon
)}}{%
1-e^{-i\pi /L(s-i\epsilon )}}
\end{eqnarray}
Hence we can compute its Laplace transform 
\begin{eqnarray}
\tilde{\mu }(z) &=&\frac{-i\lambda^{2}}{2\pi L}\int_{0}^{\infty
}\frac{dse^{-sz}}{%
s-i\epsilon }\frac{1+e^{-i\pi /L(s-i\epsilon )}}{1-e^{-i\pi /L(s-i\epsilon
)}%
}  \nonumber \\
&=&\frac{-i\lambda^{2}}{2\pi L}J(-i\epsilon z,\frac{i\pi }{2Lz})
\end{eqnarray}
where $J(x,a)$ is defined by 
\begin{equation}
J(x,a)=\int_{x}^{\infty }\frac{dye^{-y}}{y}\coth (ay)
\end{equation}
and appears in equation (IV.20) through analytical continuation in the
complex plane. This integral can actually be computed at the limit of
vanishing  $x$ ($\epsilon \rightarrow 0$) - see
reference \cite{table}, equation 3.427.4.
\begin{eqnarray}
\tilde{\mu }(z) &=&\frac{-i\lambda^{2}}{\pi ^{2}\epsilon
}+\frac{\lambda^{2}}{\pi ^{2}}%
z\log (ie^{\gamma }\epsilon z)  \nonumber \\
&&-\frac{i\lambda^{2}}{\pi L}\left[ \log \Gamma (\frac{Lz}{i\pi
})-\frac{Lz}{i\pi }%
\log (\frac{Lz}{i\pi })+\frac{Lz}{i\pi }+\frac{1}{2}\log \frac{Lz}{2i\pi
^{2}%
}\right] +O(\epsilon )
\end{eqnarray}
Note that $\tilde{\mu }(z)$ is a sum of the term of case 2 ($L\rightarrow
\infty $) and a finite one (no dependence on $\epsilon $ ) . The
logarithm of the $\Gamma $-function gives a countable number of branch cuts
at $z=-i\frac{n}{\pi L}$, $n$ positive integer, the resonance modes of the
cavity . Again the important pole  has a negative real part.  We can again
compute 
the pole perturbatively. It lies at 
\begin{eqnarray}
z &=&-i\left( \tilde{\omega}+\frac{\lambda^{2}}{\pi ^{2}}\tilde{\omega}\log
(e^{\gamma }\epsilon \tilde{\omega})-\frac{\lambda^{2}}{\pi L}\log \Gamma
(-L%
\tilde{\omega}/\pi )\right.  \nonumber \\
&&\left. -\frac{L\tilde{\omega}}{\pi }\log (-L\tilde{\omega}/\pi )-\frac{L%
\tilde{\omega}}{\pi }-\frac{1}{2}\log (-\frac{L\tilde{\omega}}{2\pi ^{2}}%
)\right)
\end{eqnarray}
Clearly the logarithm of the gamma functions is the term out of which the
real part of the pole appears. Since the real part of the pole is negative
the branch cut is excluded from the integration contour and hence 
\begin{equation}
u(s)=e^{-i\Omega s-\Gamma s}
\end{equation}
Here $\Gamma $ gives a dissipation constant.  . In Figure 1 we give a plot of the real part 
of the pole ($-\Gamma $) as a function of the frequency $\omega $. Note that it has sharp
 maxima on the resonance points, implying persistence of coherence .

Already from the approximation (IV.23) we observe that the difference in the
renormalized frequency from the case $L \rightarrow \infty$ ($\Delta \omega$%
) is finite. Unfortunately perturbation expansion is not reliable when $%
\tilde{\omega}$ is close to the resonance frequencies (this corresponds to
negative integers arguments in the $\Gamma$-function where it diverges) and
for this regime we have not been able to get any analytic results. In Figure
2 we have plotted the dependence of $\Delta \omega = \Omega[L] - \Omega[%
\infty] $ as the frequency changes. This effect of the frequency shift for
an atom within the cavity is well known, as well as its relation to the
Casimir effect \cite{Bar}.

\begin{figure}
 \centerline{%
   \psfig{file=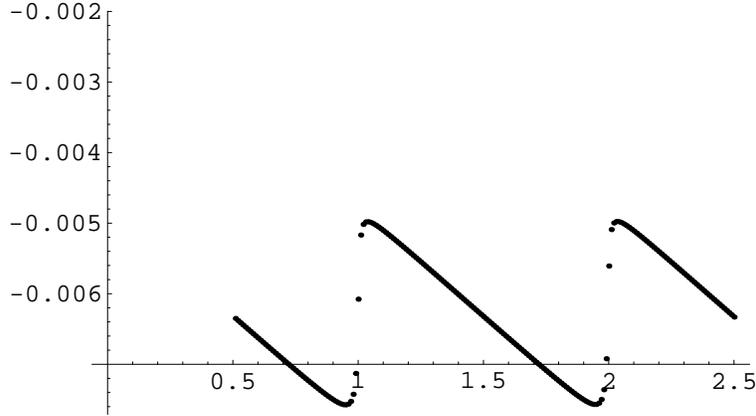,width=10cm,angle=0}%
   }
 \caption{  Real part of pole (negative dissipation constant $\Gamma$) as a function of
$\omega/(\pi/L)$. } \label{u2}
\end{figure}

\begin{figure}
 \centerline{%
   \psfig{file=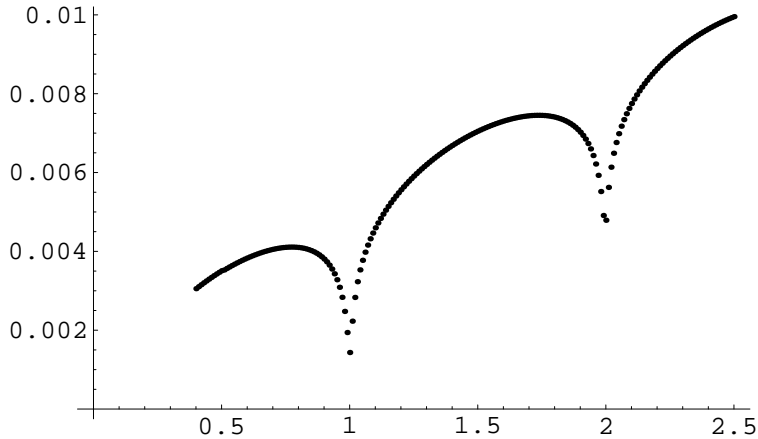,width=10cm,angle=0}%
   }
 \caption{  Frequency shift $\Delta \Omega$  as a function of
$\omega/(\pi/L)$. } \label{u3}
\end{figure}

\section{Discussion}


Let us now integrate what we have found and look at the overall picture. 
The physics of a 2LA-EMF system at zero temperature 
is characterized by a number of time constants:\\
1) The inverse natural frequency $  \omega_0^{-1}$ \\
2) The inverse coupling constant $ g_k^{-1}=\sqrt {\omega_k}/\lambda$\\
3) The relaxation time constant $  \Gamma^{-1} $\\
4) The cavity size $ L $  (divided by c)

First consider a zero temperature field in free space, thus ignoring
factors 4). Start with only one mode in the field in resonance with 
the atom, then the system undergoes  Rabi nutation with frequency $\Omega \approx g
\sqrt{n+1}$, where $n$ is the  photon number in the field.
The collapse time (assuming a large mean  photon number $\bar n$)  is  $~g^{-1} $, 
 and revival time is $~ 2 \pi \sqrt{\bar n}/g$. \cite{WM}.
Atom excitation becomes significant in a time much greater than  $
\omega_0^{-1}$ but shorter than $g^{-1}$. 
(This  is the condition for a first order perturbation 
theory to give reasonable results.)
For a large number of modes, spontaneous emission occurs at the relaxation 
time scale  $\Gamma^{-1} = \pi /g >> \omega_0^{-1}$ 
which we found  to be the same as the decoherence time -- the time for the off
diagonal elements  of the reduced density matrix to decay (Sec. 4). 
When the mean  number of photons  in the field is large ($\bar n >> 1$), 
they become comparable to the collapse time. 
This is a measure of the coherence in the atom-field system, and is
controlled mainly by their coupling and the photon number in the field.  We see that with
the resonance condition, the  nature of decoherence in 2LS   is very different from the QBM
situation, where phase information in the Brownian particle is efficiently dispersed in the many
modes in the bath coupled almost equally to the system. 
As we remarked in the Introduction, the identification of the phase information and
energy flow from the 2LS to its environment is similar to the spin echo
phenomena (Landau `damping') which is based on statistical  mixing rather than
dissipation.  The mathematical distinction lies between
considering the system coupled to the discrete 
number basis (our model) and the continuous amplitude basis (QBM) of the
environment. The latter case essentially produces noise that drives 
the system in a way insensitive to its own intrinsic dynamics. While in the
former case, the coupling respects the internal dynamical structure 
of the 2LS and allows it to keep its coherence. 

To see how the distribution of modes in a field changes the picture,  the
cavity field calculation in Sec. 4  is useful.
 As shown in Fig. 1, the relaxation constant develops peaks
and minima. The resonance effect is enhanced by a cavity size commensurate
with the natural frequency of the 2LA and dissipation  weakens. 
Narrow band resonance fluorescence as well as 
inhibition of spontaneous decay by frequent measurements -- the
Quantum Zeno effect -- are  interesting phenomena  which our equations
can provide finer details.

Non-Markovian processes involve memory effects (nonlocal in time).  
For the QBM problem, except for the case of high temperature Ohmic bath
which gives  Markovian dynamics, other types of spectral density (supraohmic)
or at low temperatures, the dynamics of the system is  non-Markovian \cite{HPZ}.
When the reaction time of the bath is comparable to or
faster than the natural time scale of the system ($\omega_0$), one also
expects to see non-Markovian behavior. By contrast, the 2LA is quite
different: At zero temperature there is 
only one timescale $\Gamma^{-1} = \lambda^{-2} \omega^{-1} >>  \omega^{-1}$ 
that determines both decoherence and relaxation. There is no memory effect
and hence the process is Markovian. We expect that in finite temperature 
the dynamics of the 2LA will  be nonMarkovian \cite{ADHS}. This is because
there are more ways for  the atom and the field to get entangled, and the
memory effects of their interaction would presumably persist.

In conclusion we find that the 2LS interacting with an EM field is far more
coherent than what is commonly believed, the  misconception probably arising
from the mistaken  identification of this system with the Brownian model
of an oscillator  interacting with a harmonic oscillator bath. 
\\ \\ \\  
{\bf Acknowledgement}
We wish to thank Phillip Johnson and Adrian Dragulescu for a close reading of our paper and making
useful suggestions, and Sanjiv Shresta for checking the formulae and  solutions to some
nonlocal differential equations numerically and providing us with the figures.
Sanjiv Shresta also first noticed a sign disrepancy in a sample calculation, which alerted us to possible 
ambiguities in the Grassmanian variable approach for finite temperature and coherent state fields.
We also thank Dr Juan Pablo Paz for pointing out a mistake in equation (III.17) of an earlier
version of this paper. This work is supported in part by NSF grant PHY98-00967.

\newpage
\begin{appendix}

\section{Atom-Field Interaction: Two-Level System}

In this Appendix we give a rather detailed derivation of the Hamiltonian
for a nonrelativistic atom interacting with a second-quantized
electromagnetic field under the dipole,  rotating wave and two level approximations.
This is to facilitate the comparison of our model (II.1)  with $\sigma_\pm$
coupling with that used by others (II.3) with $\sigma_z$ coupling. (See Introduction).
To make this also useful for later papers in this series, we have included atomic
motion.  Note that the convention here is closer to \cite{WM}
than that used in the text which is closer to \cite{VW}. The conversion is
explained in footnote 2.

The dynamics of a moving atom (mass $M$, momentum $\bP$)
whose electrons (charge $e$, mass $m$) interact with an electromagnetic
field (vector potential $\bA$, Coulomb potential $V$) is decribed by
the (classical) Hamiltonian
\be
H = \frac{\bP^2}{2M} +  \frac{1}{2m} (\bp - e \bA)^2  + e V( \bx) + H_b
\ee
where $H_b$ is the Hamiltonian for the electromagnetic field.
Expanding out one can write this as
\be
H = H_a + H_e + H_b + H_c
\ee
where
\be
H_a =\frac{ {\bP}^2 }{2M}
\ee
describes nonrelativistic atom motion,
\be
H_e = \frac{p^2}{2m} + e V(\bx)
\ee
decribes the dynamics of the (one) electron, while
\be
H_{c1} = -\frac{e}{m} {\bA} \cdot {\bp}
\ee
and
\be
H_{c2} = \frac{ e^2}{2m} {\bA}^2
\ee
describe the coupling between the electron and the field. The second
term makes no contribution to one-photon processes and will
be ignored. We will refer to
$H_0 = H_e + H_b$ as the unperturbed Hamiltonian, and
$H_I = H_{c1}$ the interaction Hamiltonian.

In a second-quantized form, the Hamiltonian for the radiation field is
given by 
\be
\hat H_b = \sum_\bk  \hbar \o_\bk  {\hat b_\bk }^\dagger {\hat b_\bk }
\ee
where ${\hat b_\bk }^\dagger$, ${\hat b_\bk }$ are the creation and annihilation
operators for the kth normal mode of a free massless vector field.
Thus for the field vacuum $ b_\bk |0\rangle = 0$, ${\hat b_\bk }|0\rangle = 0$,
 $[{\hat b_\bk },{\hat b_{\bk'}}^\dagger] = \delta_{\bk,\bk'}$,
 for all $\bk$.
We can perform a  harmonic decomposition of the vector potential of the
electromagnetic field
\be
\bA (\bx, t) = \sum_\bk  \left (\frac{\hbar}{2 \o_\bk  \e_0} \right) ^\ha
\left [b_{\bk\s} \bu_{\bk\s} (\bx) e^{i\o_\bk  t} + b^\dagger_{\bk\s} \bar \bu_{\bk\s} (\bx)
e^{-i\o_\bk  t} \right]
\ee
where, assuming the field is contained in a box of size L, the spatial
mode functions $\bu_{\bk \s}$ is given by
\be
\bu_{\bk\s} (\bx) = L^{-3/2} \hat {\bf e}_{\bk\s} f_\bk (\bx)
\ee
Here $\hat {\bf e}_{\bk\s}$ is the unit polarization vector and $\s =1, 2$ are the
two (transverse) polarizations. In free space, $f _\bk (\bx) = e^{-i \bk \cdot \bx}$.

We assume that electron motion is much faster than the motion of
the atom, thus it sees a stationary central Coulomb potential
around the center of mass of the atom.
Denoting the (time-independent, nonrelativistic) electronic wave function
eigenstate belonging to the eigenvalue $E_i$ by $\phi_i$, i.e,
$H_e\phi_i = E_i \phi_i$,
we can write the Hamiltonian for the electron in the second-quantized form as
\be
\hat H_e = \sum_i E_i \, {\hat a_i}^\dagger \, {\hat a_i}
\ee
where $i$ labels the bound states of the eletron (we assume vanishing
probability for the atom to ionize) and ${\hat a_i}^\dagger$ and ${\hat a_i}$
are the creation and annihilation operators. As fermions they obey the
anticommutation relations
${\hat a_i}^\dagger {\hat a_j} + {\hat a_j} {\hat a_i}^\dagger = \delta_{ij}$.

To perform perturbation theory, the electronic wave function of the interacting
system is expanded in terms of the eigenfunctions $\phi_i$ of the
unperturbed Hamiltonian, with basis formed by the direct product of the
electron and the field states 
Thus the electron field operator  ${\hat \psi}({\bx})$ can be expanded as
\be
{\hat \psi}(\bx) = \sum_i {\hat a}_i \phi_i(\bx).
\ee
With this, the interaction Hamiltonians is given by
\be
\hat H_{I}  = -\frac{e}{m}\int {\hat \psi}^\dagger({\bf x})\left(
   {\bf p}\cdot{\bf A} \right) {\hat \psi}({\bf x})\,d^3x
\ee
or in terms of  $\hat a_i, \hat b_\bk $ operators 
\be
\hat H_{I} = \hbar \sum_{i,j,\bk} {\hat a_i}^\dagger\, {\hat a_j}
\left(g_{ij\bk}{\hat b_\bk } + {\bar g}_{ij\bk}{\hat b_\bk }^\dagger\right)
\ee
where
\be
g_{ij\bk} = -\frac{e}{m}\frac{1}{\sqrt{2\hbar\o_\bk  \e_0}}
\int {\bar \phi}_i({\bf x}) \bu_{\bk \s} ({\bf x})\cdot {\bf p} \phi_j({\bf x})\,d^3x.
\ee
\\
\noindent {\bf Dipole Approximation}\\

Now consider conditions when the spatial variation of the vector potential $\bA$
of the electromagnetic field is small compared to the electronic
wave function $\psi$, one can expand $f_\bk(\bx)$ in $\bu_{\bk \s} (\bx)$ around the
position of the atom $ \bx = \bX +  \d \bx$:
\be
e^{i\bk \cdot \bx} = e^{i \bk \cdot \bX}[ 1 + \bk \cdot \d \bx -
\ha (\bk \cdot \d \bx )^2 + ...]
\ee
The dipole approximation amounts to keeping just the leading term.
Doing so,  we can take the field mode function $f_\bk (\bx)$ out
of the integration above and evaluate it at the atomic position. To evaluate
$$
\frac{e}{m} \int {\bar \phi_i}{\bf p}\phi_j d^3x
$$
we make use of 
\be
\frac{d \hat x_i}{dt} = \frac{\hat p_i}{m} = \frac{1}{i\hbar} [ \hat x_i, \hat H_e]
\ee
yielding, 
\be
\frac{e}{m} \int {\bar \phi_i}{\bf p}\phi_j d^3x
= i \o_{ij} \bp d_{ij}
\ee
where $\hbar\o_{ij} = E_i-E_j$, and
$\bd_{ij} \equiv e \int {\bar \phi_i}{\bf x}\phi_j d^3x$ is the dipole
matrix element, $ \bd_{ij} = {\bar \bd}_{ji}$. Define
\be
d_{ij\bk} \equiv -\frac{i\o_{ij}}{\sqrt{2\hbar\o_\bk \epsilon_0 V}}\bd_{ij}
\cdot \hat {\bf e}_{\bk \s}
\ee
Note that ${\bar d}_{ij\bk} = d_{ji\bk}$.
With this under the dipole approximation,
\be
g_{ij\bk} = d_{ij\bk} f_\bk ({\bX})
\ee
\\
\noindent {\bf Rotating Wave Approximation}\\

In the interaction picture, recalling that the time evolution of the ladder
operators are given by
\be
{\hat a_i}^\dagger (t) = {\hat a_i}^\dagger e^{i \o_it} \hspace{.5cm}
{\hat a_j}(t) = {\hat a_j} e^{-i \o_j t} \hspace{.5cm}{\rm and}\hspace{.5cm}
{\hat b_\bk }(t) = {\hat b_\bk } e^{-i \o_\bk  t}
\ee
the interaction Hamiltonian $H_{I}$ in the interaction picture becomes
\be
\tilde H_{I} =\hbar \sum_{i,j,\bk} g_{ij\bk} {\hat a_i}^\dagger\, {\hat a_j} \,
{\hat b_\bk } e^{i(\o_{ij} - \o_\bk )} 
         + \hbar\sum_{ij\bk} {\bar g}_{ij\bk} {\hat a_i}^\dagger\, {\hat a_j} \,
{\hat b_\bk }^\dagger \, e^{i(\o_{ij} + \o_\bk )}
\ee
where $\o_{ij} \equiv \o_i - \o_j$.
We see there are two types of oscillatory terms present:
$e^{-i(\o_{ij} \pm \o_\bk )}$. Processes most effective in the
absorption or emission of a photon by the atom correspond to those with
near resonance frequency $\o_{ij} \approx \o_\bk $. Assuming $\o_{ij} > 0$
($E_i > E_j$), the first type with $e^{-i(\o_{ij} + \o_\bk )}$ has a  rapidly
oscillating phase and its contribution is small compared with the second type
with  $e^{-i(\o_{ij} - \o_\bk )}$ whose stationary phase at near resonance gives a 
large contribution. Physically the first type corresponds to either the excitation
of the atom {\em along with} the emission of a photon or the relaxation
of the atom {\em along with} the absorption of a photon, which is less probable
than the second type corresponding to the excitation of an atom upon the
absorpsion of a photon or the relaxation of an atom with the emission
of a photon. We shall therefore ignore the first type of terms, which
amounts to working under the Rotating Wave Approximation. This is the second major 
approximation in this standard model.\\

\noindent {\bf Two-Level Atom}\\

Let us now consider the idealized case when the atom has only two electronic
states, $|+>, |->$ corresponding to $i=2, 1$, with energies equal to
$E_{\pm} = \pm \ha \hbar \o_0$. (The two states  can interchangeably be
labeled as  $|1>, |0>$,  or $ |e>, |g>$ or $|\uparrow>, |\downarrow>$.)  Thus 
$\o_{ij=21}= \o_0$.
Thus
\be
{\hat H}_a = \frac{\hbar \o_0}{2} (a_2^\dagger a_2 - a_1^\dagger a_1)
\equiv \frac{\hbar \o_0}{2} \s_z \equiv \hbar \o_0 S_z
\ee
where we have introduced a Pauli matrix (2x2) representation
$\s_z = diag (1, -1)$
For the interaction Hamiltonian above, under the RWA, in the $i=2, j=1$
contribution to the summation, the first line containing $e^{i(\o_0-\o_\bk )}$
is kept, while the second line is dropped. The reverse is true for the $i=1, j=2$
term. The interaction Hamiltonian (in the interaction picture) now becomes
\be
\tilde H_{I} = \hbar\sum_\bk 
\left[ g_{21\bk} {\hat a_2}^\dagger {\hat a_1} {\hat b_\bk } e^{-i(\o_\bk  - \o_{21})t}
+{\bar g}_{12\bk} {\hat a_1}^\dagger {\hat a_2} {\hat b_\bk }^\dagger
e^{-i(\o_\bk  + \o_{12})t} \right]
\ee
Introducing the Pauli matrix representation for the fermion operators
$a_2^\dagger a_1 \rightarrow \s_+  \equiv  S_+$ and
$a_1^\dagger a_2 \rightarrow \s_-  \equiv  S_-$,
 and defining $g_\bk  \equiv   d_{21\bk} = {\bar d}_{12\bk}$
 (recall $g_{ij\bk} \equiv d_{ijk} f _\bk (\bX)$)
we can write the interaction Hamiltonian (in the Heisenberg picture)
in a simple form:
\be
{\hat H}_I = \hbar \sum_\bk  g_\bk  \left( S_+ b_\bk  f_\bk (\bX) 
+ S_- b_\bk ^\dagger \bar f_\bk (\bX)) \right)
\ee
Therefore the total Hamiltonian for our model of a moving atom interacting with a
quantum electromagnetic field under the dipole, rotating wave and two-level
approximation is given by
\be
{\hat H} = \frac{{\hat {\bf P}}^2}{2M} + \hbar \o_0 {\hat S}_z
+ \hbar \sum_\bk  \left[ \o_\bk  {\hat b}_\bk ^\dagger {\hat b}_\bk 
+   g_\bk  \left( f_\bk (\bX) S_+ b_\bk   +   \bar f_\bk (\bX) S_- b_\bk ^\dagger 
\right) \right]
\ee

\section{An operator proof of the master equation}

We use the resolvent decomposition of the propagator
\begin{equation}
e^{-iHt} = \int \frac{dE  e^{-iEt}}{E - H} 
\end{equation}
Then by writing $H = H_0 + H_I$ we can expand the resolvent
and get
\begin{eqnarray}
(E - H)^{-1} = (E - H_0)^{-1} (1 - (E - H_0)^{-1}H_I)^{-1} \nonumber \\
= (E - H_0)^{-1} (1 + (E-H_0)^{-1} H_I + \frac{1}{2} (E-H_0)^{-1} H_I(E-H_0)^{-1} H_I 
+ \ldots ) 
\end{eqnarray}
When we act the expanded resolvent in any vector
 $|0, i \rangle = |0 \rangle \otimes |i \rangle$, ($|i \rangle$ denotes
an eigenstate of the Hamiltonian of the two level system) we see that \\
 only expanded terms that contain alternating sequences of $\sigma_+$ and
$\sigma_-$ survive. This makes the summation much easier.
 Also if we note that 
\begin{eqnarray}
(E - H_0)^{-1} \sum_k g_k b_k  (E - H_0)^{-1} g_{k'} \sum_{k'} b^{\dagger}_{k'} |0 \rangle \nonumber \\
 = (E - \omega_0)^{-1} \sum_k \frac{g_k^2}{E - \omega_k}  | 0 \rangle := 
( E - \omega_0)^{-1} F(E)
\end{eqnarray}
Hence we can compute the matrix elements by resumming the expansion
\begin{eqnarray}
\langle z ,0|(E - H)^{-1}|0,0  \rangle &=& E^{-1} \\
\langle z , 1 |(E-H)^{-1} | 0,0 \rangle &=& 0 \\
\langle z,1|(E - H)^{-1} | 0,1 \rangle &=& (E - \omega_0 - F(E))^{-1} \\
\langle z,0 | (E - H)^{-1} |0,0 \rangle &=& \frac{E - \omega_0}{E - \omega_0 - F(E)} 
\sum_k \frac{g_k}{(E - \omega)^2} \bar{z}_k
\end{eqnarray}

The reduced density matrix  propagator in the energy basis of the two- 
level atom and for the vacuum initial state is 
\begin{eqnarray}
J(i,j;t|m,n;0) = \int dE dE' e^{-i(E-E')t} \int Dz D\bar{z} e^{- \sum_k \bar{z}_k z_k} \\ \nonumber
 \left( \langle z, i|(E - H)^{-1} |0,m  \rangle \langle 0, n| (E-H)^{-1}|z,j \rangle
\right)
\end{eqnarray}
\par
We can then verify that the only nonzero elements $mn \rightarrow ij$ are 
the following and their conjugates 
\begin{eqnarray}
J(0,0; t|0,0;0) \rightarrow  E^{-1} E'^{-1} \\
J(0,1;t| 0,1;0) \rightarrow  E^{-1} (E' - \omega_0 - F(E'))^{-1} \\
J(0,0;t|0,0;0)  \rightarrow  \frac{E - \omega_0}{E - \omega_0 - F(E)} 
\frac{E' - \omega_0}{E' - \omega - F(E')} \sum_k \frac{g_k^2}{(E- \omega_k)
(E' - \omega_k)} \\
J(11;t|1,1;0) \rightarrow  (E - \omega_0 - F(E))^{-1} (E - \omega_0 - F(E'))^{-1} 
\end{eqnarray}
Then it is easy to check that this reproduces the propagation as given 
by equation (3.21) that was obtained through  the influence functional
method. Indeed 
\begin{equation}
u(t) = \int \frac{dE e^{-iEt}}{E - \omega_0 - F(E)} 
\end{equation}
is exactly the same as the one defined by equation (3.9). 
\end{appendix}

\newpage

\end{document}